\documentclass[
 sort&compress
    ,final            % use final for the camera ready runs
  ]
  {aipproc}

\usepackage{epsfig}
\layoutstyle{8x11double}

\begin{document}

\newcommand{\lsim}{\raisebox{-0.13cm}{~\shortstack{$<$ \\[-0.07cm] $\sim$}}~}
\newcommand{\gsim}{\raisebox{-0.13cm}{~\shortstack{$>$ \\[-0.07cm] $\sim$}}~}

\title{(SUSY) Higgs Search at the LHC}


\author{M.~Margarete M\"uhlleitner}{
  address={LAPTH, 9 Chemin de Bellevue, B.P.110, 74941 Annecy-Le-Vieux, Cedex, France}
}

\begin{abstract}
The discovery of the Standard Model (SM) or supersymmetric (SUSY) Higgs 
bosons belongs to the main endeavors of the Large Hadron Collider (LHC). 
In this article the status of the signal and background calculations for 
Higgs boson production at the LHC is reviewed.
\end{abstract}

\maketitle

%%%%%%%%%%%%%%%%%%%%%%%%%%%%%%%%%%%%%%%%%%%%
%% MAINMATTER
%%%%%%%%%%%%%%%%%%%%%%%%%%%%%%%%%%%%%%%%%%%%

\section{Introduction}

The Higgs mechanism is a cornerstone of the SM and its SUSY extensions 
\cite{higgsmech} and renders the search for Higgs boson(s) a crucial endeavor 
at the LHC. In the SM one Higgs doublet has to be introduced to realize 
electroweak symmetry breaking (EWSB), leading to one elementary Higgs boson 
\cite{higgs}. Experiment and theory constrain the Higgs boson mass range to 
114.4~GeV up to $\sim 1$~TeV \cite{constraints}. The minimal supersymmetric 
extension of the SM (MSSM) requires at least two complex Higgs doublets, 
implying 5 physical Higgs states after EWSB, two neutral CP-even, $h,H$, one 
neutral CP-odd $A$ and two charged Higgs bosons $H^\pm$. Negative direct 
searches at LEP2 impose lower mass bounds of $m_{h,H} \gsim 92.8$~GeV, 
$m_A \gsim 93.4$~GeV and $m_{H^\pm} \gsim 78.6$~GeV \cite{mssmconstraints}. 
SUSY implies an upper light neutral Higgs mass bound of $m_h \lsim 135$~GeV
\cite{hmasslim}, including the one-loop and dominant two-loop corrections. The 
heavier Higgs boson masses range up to ${\cal O}(1$~TeV). At tree level, the 
MSSM Higgs sector can be parametrized by two independent input parameters, $m_A$
and $\mbox{tg}\beta=v_2/v_1$, the ratio of the two vacuum expectation values. 

\section{Gluon Fusion}

\begin{figure}
\epsfig{figure=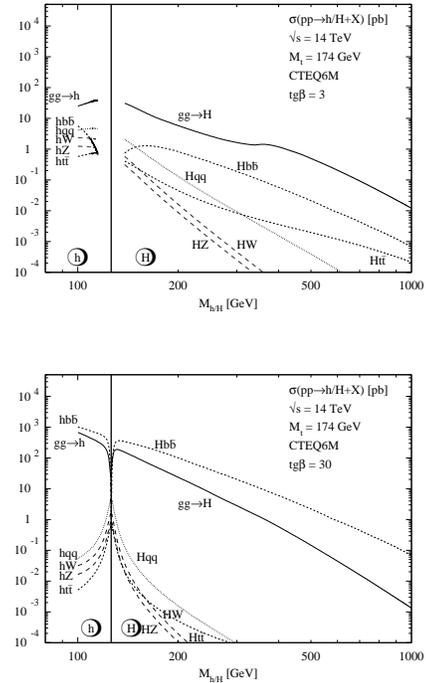,bbllx=43,bblly=115,bburx=546,bbury=756,angle=-90,width=5.5cm,clip=}
\caption{Neutral CP-even MSSM Higgs production cross sections at the LHC for
gluon fusion ($gg\to h/H$), vector boson fusion ($qqh/H$), 
Higgs-strahlung ($hV/HV$), and the associated production 
($b\bar{b}h/H,t\bar{t}h/H$), including QCD corrections except for 
$b\bar{b}h/H$. Upper: $\mbox{tg}\beta=3$, lower: $\mbox{tg}\beta=30$.
From Ref.~\cite{higgsmech}.}
\label{mssmprod}
\end{figure}

\begin{figure}[h]
\epsfig{figure=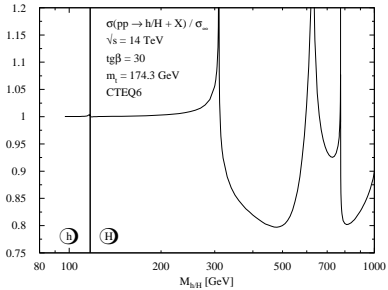,bbllx=48,bblly=222,bburx=568,bbury=628,width=5cm,clip=}
\caption{Ratio of the QCD corrected scalar MSSM Higgs production cross
sections via
gluon fusion including the full squark mass dependence and those obtained by
taking the relative QCD corrections to the squark loops in the heavy mass limit
as functions of the corresponding Higgs masses for $\mbox{tg}\beta=30$.
The kinks and spikes correspond to the various squark pair thresholds.
From Ref.~\cite{qcdsqggfus}.}
\label{relsquark}
\end{figure}

The gluon fusion processes 
\begin{eqnarray}
gg\to \Phi \qquad (\Phi=H^{\bf{SM}},h,H,A)
\end{eqnarray}
are the dominant production mechanisms for the SM
Higgs boson in the entire mass range up to $\sim 1$~TeV and for MSSM Higgs
states for small to moderate values of $\mbox{tg}\beta$ 
(see Fig.\ref{mssmprod}). It is 
mediated by heavy quark triangle loops, and in addition by squark loops in 
SUSY theories, if $m_{\tilde{q}} \lsim 400$~GeV. 
%%%%%%%%%%%%%%%%%%%%%%%%%%%%%%%%%%%%%%%%%%%%%%%%%%%%%%%%%%%%%%%%%%%%%%%%%%%
\begin{figure}[h]
\epsfig{figure=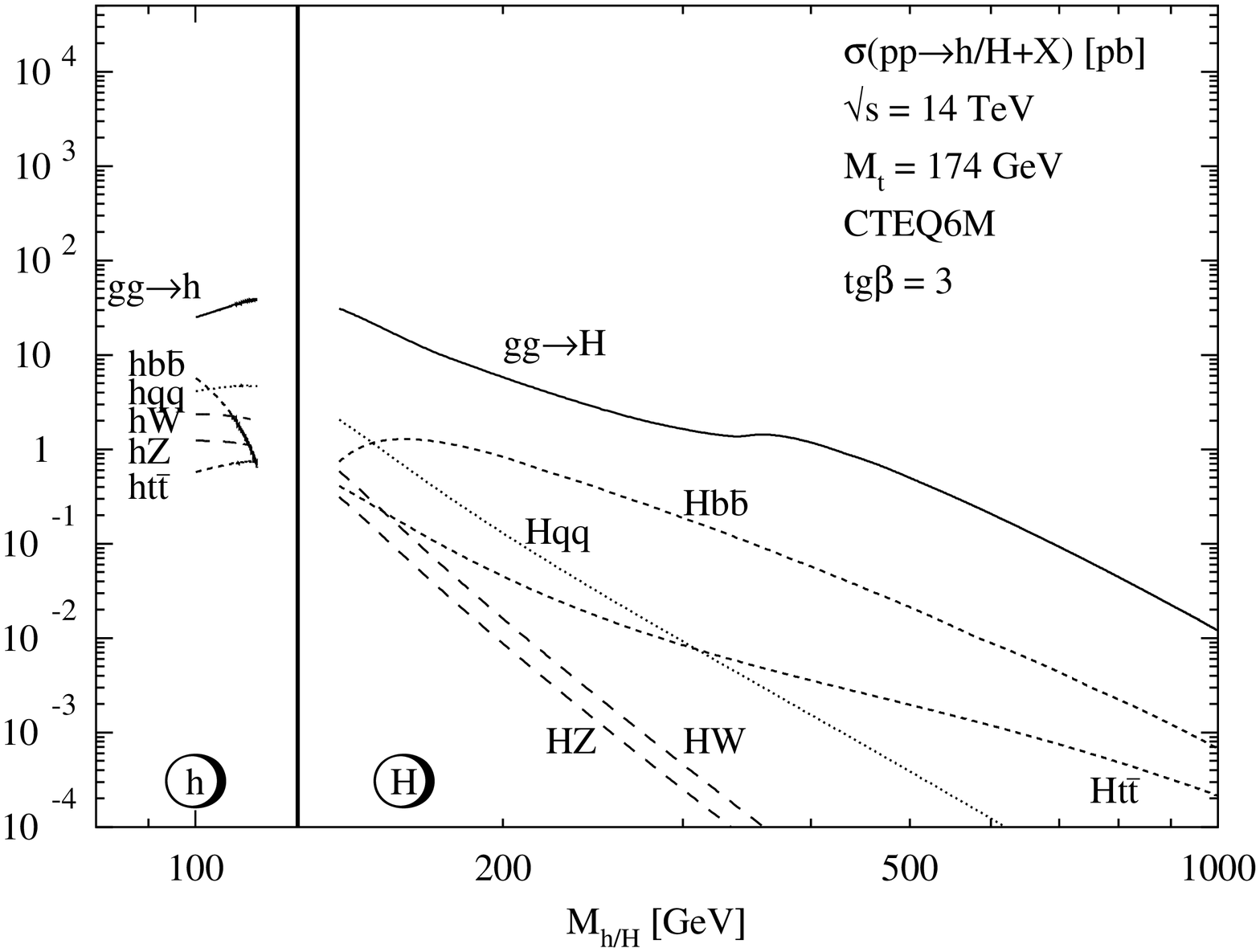,bbllx=43,bblly=115,bburx=546,bbury=756,angle=-90,width=5.5cm,clip=} 
\end{figure}
%%%%%%%%%%%%%%%%%%%%%%%%%%%%%%%%%%%%%%%%%%%%%%%%%%%%%%%%%%%%%%%%%%%%%%%%%%%
The next-to-leading order
(NLO) QCD corrections, including the full quark
mass dependence, increase the production cross section by up to 100\% 
\cite{nloggfus}. The calculation in the heavy top quark 
limit \cite{ggfustoplim}
has been shown to provide an approximation within 10~\% in the SM and 
20-30\% in the MSSM for $\mbox{tg}\beta\lsim 5$ \cite{approxconfirm}.
For large $\mbox{tg}\beta$ the bottom loop contributions become dominant 
due to the strongly enhanced bottom Yukawa couplings, so that the large 
loop mass limit is not applicable any more. In this case, the NLO 
corrections are of more moderate size, 
${\cal O}(10-60$\%).
The next-to-next-to leading order (NNLO) corrections, provided in the
heavy top limit, increase the cross section moderately by another 
20-30\% \cite{nnloggfus}. 
The effect of the dominant finite top mass corrections on the
approximate NNLO result has been investigated recently in Ref.~\cite{marzani}.
Soft gluon resummation leads to a further increase of 
about 6\% \cite{gluonresum} and the estimate of the next-to-next-to-next-to
leading order (NNNLO) effects indicates improved perturbative convergence
\cite{nnnlo}. 
%These results can also be applied to the scalar MSSM Higgs cross 
%sections in the regions where the top loops are dominant, {\it i.e.} for
%small values of $\mbox{tg}\beta$.
Two-loop electroweak
(EW) corrections add $\sim 5-8$\% \cite{ewcorrggfus}.
As for the MSSM processes, the NLO QCD corrections to squark loops have 
been first known in the heavy squark limit \cite{squarklimggfus}, and 
the full SUSY-QCD corrections in the heavy SUSY particle mass limit
\cite{susyqcdlimggfus}. They are large for the squark loops,
whereas the genuine SUSY-QCD corrections, mediated by virtual gluino and stop
exchanges, are small, of ${\cal O}(5$\%).
The calculation of the QCD corrections to squark loops including
the full mass dependence has been performed recently indicating that
the squark mass effects on the $K$-factor, describing the ratio between the
NLO and LO cross section, can be up to $\sim 20$\% with a remaining residual 
theoretical uncertainty of less than about 20\% \cite{qcdsqggfus}, see 
Fig.~\ref{relsquark}. (The virtual corrections have been derived in 
Ref.~\cite{virtsquark}.)
Recently, the NLO SUSY-QCD corrections including the full SUSY particle mass 
dependences have been calculated \cite{fullsqcdggfus}.
%%%%%%%%%%%%%%%%%%%%%%%%%%%%%%%%%%%%%%%%%%%%%%%%%%%%%%%%%%%%%%%%%%%%%%%%%%%%%%%
%%%%%%%%%%%%%%%%%%%%%%%%%%%%%%%%%%%%%%%%%%%%%%%%%%%%%%%%%%%%%%%%%%%%%%%%%%%%%%%

The impact of the higher order (HO) 
calculations on the rate and the shape of the 
corresponding distributions may strongly depend on the choice of cuts. This
has therefore to be studied to perform reliable experimental analyses. 
The transverse momentum $p_T$ distribution of the Higgs boson plays an 
important role for the experimental search strategies. At LO Higgs bosons are 
produced with vanishing transverse momentum in the gluon fusion process. 
They acquire non-zero $p_T$ only if an additional gluon is radiated. This 
contribution is part of the 
real NLO corrections to the total gluon fusion cross section.
The $p_T$ distributions are known up to NLO in the heavy quark
limit \cite{gg1jnlo}, so that they are valid only for small and moderate Higgs
masses and $p_T$. For $p_T$ values much smaller than $M_H$ large logarithms
appear which necessitate resummation in order to get a reliable description
of the small $p_T$ range. This has been done at different levels of 
theoretical accuracy \cite{resumpt}. 
%Monte Carlo even generators approximatively perform
%the $p_T$ resummation. For a comparison of the different tools see 
%Ref.\cite{comparison}.
Finally, the NLO corrections with two tagged jets at large $p_T$ are 
known in the large $m_t$ limit \cite{gg2jnlo} and have 
been implemented in parton level Monte Carlo programs. 

For the MSSM, the Higgs plus 1 jet process is known at LO only 
\cite{mssmgg1j,langenegger}.
The SM results at NLO QCD for the $p_T$ distributions and the resummation
of the soft gluon effects can only be applied for small values of 
$\mbox{tg}\beta, M_H$ and $p_T$. For large $\mbox{tg}\beta$ values, the bottom
loops are dominant and as an important consequence the $p_T$ distributions
of the neutral Higgs bosons are softer than for small values of 
$\mbox{tg}\beta$ \cite{langenegger}.

As for the NNLO accuracy the inclusive SM cross section with a jet veto was 
provided some time ago \cite{nnlojetveto}. In the meantime the NNLO result
fully taking into account experimental cuts has been performed in the $H\to
\gamma\gamma$ \cite{nnlohgamgam} and in the $H\to W^+W^-\to l^+ l^- \nu
\bar{\nu}$ \cite{nnlohww} decay mode. An independent calculation for these
and the $H\to ZZ\to 4$ leptons decay channel has been finished recently 
\cite{indep} and implemented in a Monte Carlo program. 

\vspace*{0.1cm}
\underline{\it Search Channels \& Backgrounds:} 
Higgs particles with masses below $\sim 140$~GeV have to be
searched for in their $\gamma\gamma$ decay channel, the $b\bar{b}$ final 
state being overwhelmed by the QCD background. 
The $\gamma\gamma$
irreducible background has been calculated up to NLO including fragmentation
effects \cite{binoth}. The $gg\to\gamma\gamma$ contribution, which is enhanced
by the large gluon luminosity, is known up to NLO \cite{bern}. 

For SM Higgs masses $140\lsim M_H \lsim 180$~GeV the $H \to W^+ W^- \to
l^+ l^- \nu\bar{\nu}$ decay mode is among the most promising for an early 
discovery. Challenging due to the large background, the strong 
angular correlations between the charged leptons from
the $W$ decay \cite{dittmar} can be exploited to suppress the background.
Since the Higgs mass cannot be reconstructed directly due to the escaping 
neutrinos a background extrapolation from sidebands is impossible. 
It has to be 
extrapolated from the signal-free region and therefore requires a precise 
knowledge of the background distributions. The NLO $W^+W^-$ irreducible 
background is known \cite{nlowwbkg} including spin correlations, and the 
effects of multiple soft-gluon emissions up to next-to-leading-log (NLL) 
accuracy \cite{grazznll}. The knowledge of spin correlations is crucial 
for a correct prediction of angular distributions, and they are implemented
in the MC@NLO event generator \cite{spincorrwwbkg}.
Also the potentially large gluon initiated contribution $gg\to W^+W^-$ 
is available at LO \cite{ggww}. Recently the NLO QCD corrections to $W^+
W^-j$ production, an important background also to the $WW$ fusion Higgs boson
production process (see below), has been reported \cite{wwjet}.
The important $t\bar{t}$ background is known
at NLO \cite{nlottbkg}, including the effect of spin 
correlations \cite{ttbkg}. The  inclusion of width effects is 
available at LO only \cite{kauer}. 
%
%The efficiency of experimental cuts in various approximations of higher order
%effects has been studied recently for the $pp\to H\to WW$ signal cross 
%section \cite{anastasiou}.

The gold-plated channel $H\to ZZ \to 4$~leptons becomes dominant for SM Higgs
bosons with $M_H \gsim 180$~GeV. Since the invariant mass of the leptons can be 
reconstructed,
it is much easier to observe, and the background is measurable from the data.
To study the Higgs boson properties, an accurate prediction is necessary, 
however. The irreducible $ZZ$ background has been provided up to NLO including
spin correlations \cite{nlowwbkg}, the impact of soft gluon effects on signal
and background has been examined \cite{softgluon}, and the gluon-initiated 
contribution $gg\to ZZ$ is available at LO \cite{ggzz}. 
The full QCD+EW corrections to the $H\to W^+ W^- (ZZ) \to 4$~leptons decay 
have been computed by Ref.~\cite{hdeccorr}.

\section{Vector boson fusion}

The $W$ and $Z$ boson fusion processes \cite{vbf}
\begin{eqnarray}
pp \to qq \to qq + WW/ZZ \to qq H
\end{eqnarray}
for SM Higgs boson production are about one order of magnitude smaller than 
the gluon fusion process and become competitive with the latter for large
Higgs masses. The typical signature 
of a vector boson fusion (VBF) event is given by two hard jets with a 
large rapidity interval between them. Since the exchanged boson is 
colourless, there is no hadronic activity between them, and the NLO QCD 
corrections can be derived from the NLO corrections to deep inelastic
lepton-nucleon scattering. They are of ${\cal O}(10$\%), neglecting small
interference effects \cite{nlovbf}. 
Recently, the full EW and QCD corrections 
have been computed with a typical size of $5-10$\%. 
They induce distortions of the distributions at the 10\% level \cite{ewvbf}.  
%The NLO QCD corrections have been also implemented for the distributions 
%\cite{distrvbf}. 

In the MSSM, the VBF processes play an important role for the light scalar
Higgs $h$ close to its upper mass bound, where it becomes SM-like, and for
the heavy scalar Higgs $H$ at its lower mass bound. In the other regions the 
cross sections are suppressed due to additional SUSY-factors in the Higgs
couplings. For the pseudoscalar $A$ the process is absent, since it does not 
couple to gauge bosons at tree level. 
The NLO QCD corrections to the total cross sections
and distributions can be taken from the SM case and are of the same size.
The SUSY-QCD \cite{sqcdvbf} and SUSY EW and QCD corrections 
\cite{sqcdewvbf} turned out to be small. 

\vspace*{0.1cm}
\underline{\it Search Channels \& Backgrounds:} 
The Higgs $+2$ jets signature, also produced in gluon fusion, is part of the 
inclusive Higgs signal, but represents a background to the isolation of the 
Higgs gauge couplings through VBF. The gluon contribution is known at LO with 
full top mass dependence \cite{gg2jlo}, and has been recently complemented by 
the NLO QCD corrections in the heavy top mass limit \cite{gg2jnlo}. Parton 
shower effects on the relevant distributions have been evaluated as well 
\cite{parton} and indicate that the discriminating power 
of the LO results is not significantly changed.
Furthermore, the dominant 
NLO QCD corrections to Higgs plus 3jet production in vector boson fusion
have been presented in \cite{vbf3j}. The corrections to the total cross 
sections are modest, while the shapes of some kinematical distributions 
change appreciably. 

The Higgs decays into $\tau$ lepton pairs constitute an important MSSM Higgs 
boson discovery channel \cite{mssmtautau}. 
Since in VBF Higgs bosons are produced at large transverse 
momentum, the $\tau^+\tau^-$ invariant mass can be reconstructed with an 
accuracy of a 
few GeV, and the background can in principle be measured from the data using a
sideband analysis. The most important backgrounds are QCD $Zjj$ and EW $Zjj$ 
production from VBF. They are known at NLO accuracy \cite{nlozjj}.

The $H\to W^+W^- \to l^+l^- \nu\bar{\nu}$ decay, important for SM Higgs 
particle searches in the intermediate mass range \cite{zeppenfeld}, is most 
challenging, because it does not allow a direct Higgs mass reconstruction. 
The irreducible $W^+W^-$ background is calculated up to NLO, the corrections
are modest \cite{wwtowwnlo}. The other important background $t\bar{t}+$ jet
is known at NLO \cite{ttjnlo}. The essential inclusion of the top decay with 
full spin correlation is missing so far, as well as the possibly relevant 
finite width effects.

\section{Associated production with top quarks}

Associated Higgs production with a top quark pair \cite{assoctth}
\begin{eqnarray}
pp \to q\bar{q}/gg \to \Phi t\bar{t}
\end{eqnarray}
plays a significant role for SM Higgs masses below $\sim 150$~GeV and in the MSSM
only for the light scalar Higgs state $h$. The NLO QCD corrections have been
determined \cite{nlotth} and lead to a moderate increase of the total cross 
section by $\sim 20$\% at the LHC, both 
for the SM and the MSSM case. The relevant parts of 
final state particle distribution shapes are only moderately affected, i.e.
${\cal O}(10$\%). The SUSY-QCD corrections to $t\bar{t}h$ production
have been calculated in \cite{peng,hollikrauch} and are of moderate 
size.

\vspace*{0.1cm}
\underline{\it Search Channels \& Backgrounds:} 
While the $t\bar{t}H$ production with subsequent $H\to b\bar{b}$ decay was 
previously considered a discovery channel, recent analyses with a more careful
background evaluation are not as optimistic and demand for improved signal and
background studies based on NLO calculations to reliably analyse the potential 
of this channel. After the NLO signal \cite{nlotth} and $t\bar{t}j$ \cite{ttjnlo}
predictions, recently the first step towards the full NLO QCD $pp \to t\bar{t} b
\bar{b} +X$ calculation has been accomplished by evaluating the contribution
from quark-antiquark annihilation \cite{nlottbb}.
The $t\bar{t}H \to t\bar{t} \gamma\gamma$ channel on the other hand provides an 
important contribution to Higgs discovery in the $\gamma\gamma$ final state 
\cite{cmsatlas}. It 
develops a narrow resonance in the invariant $\gamma\gamma$ mass distribution,
so that the background can be measured directly from the sidebands.

\section{Associated production with bottom quarks}

While unimportant in the SM, in the MSSM for large values of $\mbox{tg}\beta$
the Higgs boson production in association with bottom quarks
\begin{eqnarray}
pp \to q\bar{q}/gg \to \Phi b\bar{b} \quad \Phi=h,H,A
\end{eqnarray}
constitutes the dominant Higgs production process. Two different formalisms,
which represent different orderings of perturbation theory, can be applied
to calculate the cross section. In the four-flavour scheme, the LO process
is given by $gg\to b\bar{b}\Phi$. The NLO QCD corrections can be deduced from
the analogous calculation with a top quark final state pair and turn out
to be large \cite{nlobbh}. This may be due to the
large logarithms generated by the integration over the transverse momenta of
the final state bottom quarks. 
The NLO corrections with one or two tagged high-$p_T$ $b$-jets have been 
provided as well 
\cite{nlobbh}. To resum the logarithms bottom 
quark densities have to be introduced in the proton. In the five flavour
scheme the LO process is then given by $b\bar{b} \to \Phi$. 
The NLO \cite{nlobbphi} and NNLO \cite{nnlobbphi} QCD corrections to the 
%%%%%%%%%%%%%%%%%%%%%%%%%%%%%%%%%%%%%%%%%%%%%%%%%%%%%%%%%%%%%%%%%%%%%%%%%%%%%%
\begin{figure}[h]
\epsfig{figure=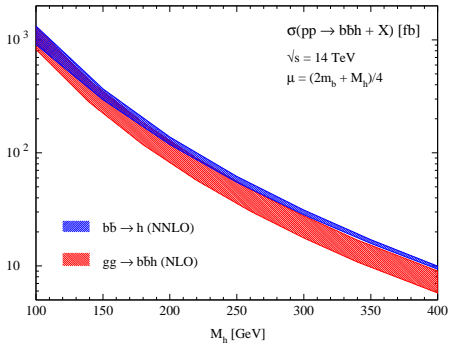,bbllx=48,bblly=222,bburx=568,bbury=628,width=6cm,clip=}
\caption{Total cross sections $pp\to b\bar{b}h+X$ as a function of the Higgs mass $M_h$ with no $b$-jet identified in the final state. The error bands correspond to varying the renormalization and factorization scale within $(2m_b+M_h)/8$ and $(2m_b+M_h)/2$. From Ref.\cite{detailsbbphi}.}
\label{bbhfig}
\end{figure}
%%%%%%%%%%%%%%%%%%%%%%%%%%%%%%%%%%%%%%%%%%%%%%%%%%%%%%%%%%%%%%%%%%%%%%%%%%%%%%
inclusive process are known and of moderate 
size. For the process with one 
tagged high-$p_T$ $b$-jet they have been calculated at NLO 
\cite{nlobbphi1tag}. At sufficiently high order in perturbation theory, 
the results of the two approaches should converge against the same value, 
see Fig.\ref{bbhfig} and Ref.~\cite{detailsbbphi} for details. 

Furthermore, the EW corrections to 
$b\bar{b} \to \Phi$ \cite{muck} and the Yukawa coupling induced 
contributions to $gg\to b\bar{b} h$ at one loop \cite{ninh} have been calculated.
The SUSY-QCD corrections to $b\bar{b}h$ have been given in 
\cite{hollikrauch} and turn out to be significant. 
(Approximative results have been given in Ref.~\cite{gao}.)
The SUSY
QCD corrections to $\Phi b$ production \cite{susyqcdbbphi} are typically of
order of a few percent after appropriate redefinition of the Yukawa couplings.

\section{Higgs-strahlung off vector bosons}
For SM Higgs masses in the intermediate range $M_H \lsim 2 M_Z$ the 
Higgs-strahlung off $W,Z$ bosons \cite{higgsrad}
\begin{eqnarray}
pp \to q\bar{q} \to Z^*/W^* \to H + Z/W
\end{eqnarray}
provides alternative Higgs boson search signatures. For the neutral MSSM
Higgs bosons, however, this process plays no major role at the LHC. The NLO
QCD corrections can be inferred from the Drell-Yan process and increase the 
total cross section by ${\cal O}(30$\%) \cite{nlohrad}. The NNLO corrections 
\cite{nnlohrad} lead to an increase of less than 10\% for the relevant Higgs 
boson masses at the LHC. The full EW corrections decrease the total cross 
section by 5-10\% \cite{ewhrad}. The total theoretical uncertainty is of 
${\cal O}(5$\%). The NLO EW and NNLO QCD corrections are similar in the MSSM 
case. The SUSY-QCD corrections are small \cite{sqcdvbf}, whereas the SUSY-EW 
corrections are unknown.

\section{Charged Higgs boson production}
The dominant charged Higgs boson production process is the associated 
production with heavy quarks \cite{assocchar}
\begin{eqnarray}
pp \to q\bar{q},gg \to H^- + t\bar{b} \qquad \mbox{and } \quad c.c. 
\end{eqnarray}
The NLO QCD and SUSY-QCD corrections are significant \cite{nlochar},
partly due to logarithms which arise from the integration over the 
transverse momentum of the final state bottom quark, and due to the large
SUSY-QCD corrections to the bottom Yukawa coupling. In analogy to the 
neutral Higgs boson case, the logarithms can be resummed by introducing
bottom quark densities. In this 5-flavour approach the LO process is then
given by $gb \to H^-t$ and its charge conjugate. The NLO SUSY-QCD corrections
to this process have been found to be significant \cite{nlogbchar}. The
next important production process is provided by the Drell-Yan type charged
Higgs pair production in 
\begin{eqnarray}
pp \to q\bar{q} \to H^+ H^- \;,
\end{eqnarray}
mediated by s-channel $\gamma$ and $Z$ boson exchange. The NLO QCD corrections
which can be inferred from the Drell-Yan process are of moderate size. The
SUSY-QCD corrections are mediated by virtual gluino and squark exchange in
the initial state and are small \cite{sqcdvbf}. Another source of 
charged Higgs pair production is the loop-mediated process from $gg$ initial
states \cite{ggchar,bbchar}
\begin{eqnarray}
pp \to gg \to H^+ H^- \;,
\end{eqnarray}
known at LO only. It is of similar size as the bottom-initiated process 
\cite{bbchar}
\begin{eqnarray}
pp \to b\bar{b} \to H^+ H^- \;,
\end{eqnarray}
which relies on the introduction of bottom quark parton densities. The NLO
corrections have been computed \cite{sqcdbbchar},
and the SUSY-QCD corrections are of significant size. The pure QCD and 
genuine SUSY-QCD corrections can be of opposite sign.
Another production mechanism is in 
association with a $W$ boson \cite{charassocw,charassocw1}
\begin{eqnarray}
pp \to gg \to H^+ W^- \qquad \mbox{and } \quad c.c.
\end{eqnarray}
The process, which is known at LO only, is mediated by top-bottom loops and, if 
their masses are light enough, by stop-sbottom loops. The corresponding 
bottom-quark initiated process in the five flavour approximation is 
\cite{charassocw,bbcharassocw}
\begin{eqnarray}
pp \to b\bar{b} \to H^+ W^- \qquad \mbox{and } \quad c.c.
\end{eqnarray}
The QCD are known and are of moderate size \cite{qcdbbcharassocw}.

\section{Conclusions}
In summary, the important Higgs boson production cross sections at the LHC are 
theoretically under good control. Most (SUSY) QCD corrections are known.
The corrections are large in several cases, and the remaining theoretical 
uncertainties have decreased from LO 100\% down to $\lsim 15$\%. Considerable 
progress has also been made in the calculation of the dominant background 
processes up to NLO. Finally, there are a lot of programs available including 
many of these corrections. 

%%%%%%%%%%%%%%%%%%%%%%%%%%%%%%%%%%%%%%%%%%%%%%%%
%% BACKMATTER
%%%%%%%%%%%%%%%%%%%%%%%%%%%%%%%%%%%%%%%%%%%%%%%%

\begin{theacknowledgments}
I would like to thank the organizers of SUSY08 for their support and 
hospitality.
\end{theacknowledgments}

\bibliographystyle{aipproc}   % if natbib is available
%\bibliographystyle{aipprocl} % if natbib is missing

%%%%%%%%%%%%%%%%%%%%%%%%%%%%%%%%%%%%%%%%%%%
%% Bibliography
%%%%%%%%%%%%%%%%%%%%%%%%%%%%%%%%%%%%%%%%%%%
%\bibliography{sample}

%%%%%%%%%%%%%%%%%%%%%%%%%%%%%%%%%%%%%%%%%%%
%% Just a reminder that you may have to run bibtex
%% All of it up to \end{document} can be removed
%% if you don't like the warning.
%%%%%%%%%%%%%%%%%%%%%%%%%%%%%%%%%%%%%%%%%%%
%\IfFileExists{\jobname.bbl}{}
% {\typeout{}
%  \typeout{******************************************}
%  \typeout{** Please run "bibtex \jobname" to optain}
%  \typeout{** the bibliography and then re-run LaTeX}
%  \typeout{** twice to fix the references!}
%  \typeout{******************************************}
%  \typeout{}
% }

\end{document}